\documentclass{ws-p9-75x6-50}

\begin{document}

\title{After MAP: Next Generation CMB}

\author{Asantha Cooray}

\address{Theoretical Astrophysics, California Institute of Technology,
Pasadena, CA 91125\\E-mail: asante@caltech.edu}

\maketitle

\abstracts{We discuss several opportunities involving
cosmic microwave background (CMB) observations during the post-MAP era.
The curl-modes of CMB polarization allow a direct detection of inflationary gravitational waves
and a measurement of the energy scale of inflation. While a significant source of confusion is expected from
cosmic shear conversion of polarization related to density perturbations,
higher resolution observations of CMB anisotropies can be used
for a lensing reconstruction and to separate gravitational-wave polarization signature from that of lensing.
With perfect all-sky maps, separations based on current lensing reconstruction techniques allow
the possibility to probe inflationary energy scales down to 10$^{15}$ GeV, well below that of grand unified theories.
Another aspect of future CMB studies will be related to large scale structure, such as wide-field imaging
of Sunyaev-Zel'dovich (SZ) effect in galaxy clusters. Here, we comment
on a potentially interesting and unique application of the SZ effect
involving a measurement of the absolute temperature of CMB  at high redshifts.
Finally, we comment on cluster polarization signals and the possibility to directly measure some aspect
of the temperature quadrupole associated with time evolving potentials.}

Observations of the cosmic microwave background (CMB)  temperature fluctuations 
have now established the presence of now well-known acoustic peak structure 
in the anisotropy power spectrum\cite{Miletal99}. Along with constraints on cosmological parameters, these
observations provide evidence for an initial spectrum of scale-invariant adiabatic density perturbations;
Initial quantum fluctuations associated with an inflationary origin may be responsible here.
While this is not a direct confirmation of inflation, 
it has been argued for a while that the smoking-gun signature of inflation would be the detection of 
stochastic background of gravitational waves associated with it\cite{KamKos99}.  These 
gravitational-waves produce a distinct signature in the polarization of
CMB in the form of a contribution to the curl, or magnetic-like, component of the 
polarization\cite{KamKosSte97}. While polarization from density, or scalar, perturbations dominate,
due to the fact they have no handedness, there is no contribution to curl mode polarization from density perturbations.

\begin{figure}[t]
\includegraphics[height=15pc,angle=-90]{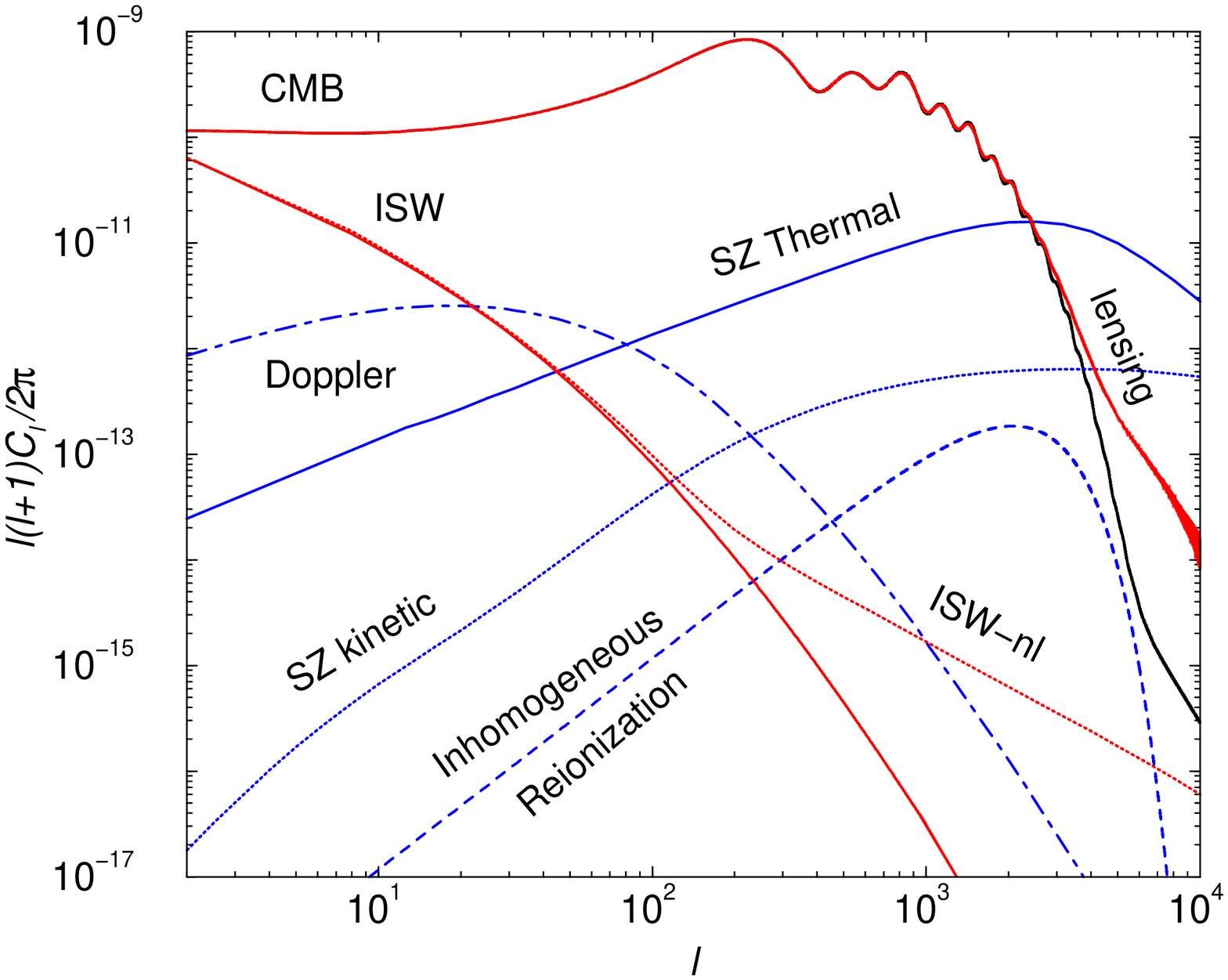} \includegraphics[height=15pc,angle=-90]{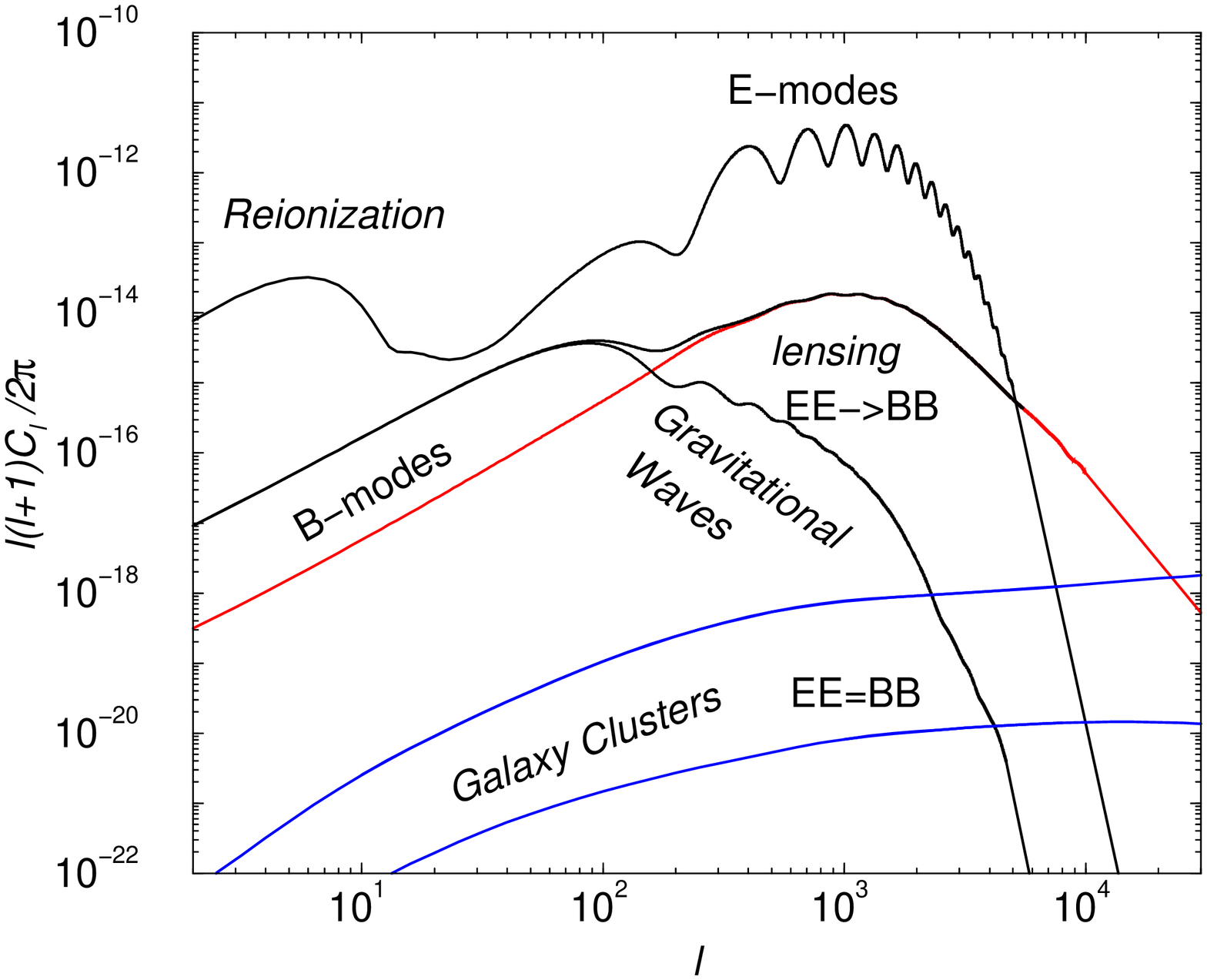}
\caption{Power spectrum for the temperature ({\it left}) and polarization
({\it right}) anisotropies in the fiducial $\Lambda$CDM model
with $\tau=0.1$. In the case of temperature, the curves show the 
local universe contributions to CMB due to
gravity (ISW and lensing) and scattering (Doppler, SZ effects, 
patchy reionization). In the case of polarization, local universe contributes
through rescattering via free electrons in galaxy clusters and
the lensing conversion of E and B-modes.}
\label{fig:sec}
\end{figure}

In Figure~\ref{fig:sec}, we show the contribution from dominant scalar modes to 
the polarization in the gradient component, or the so-called E-modes,
 and from gravitational-waves to the curl polarization (B-modes). 
We have assumed here an inflationary energy scale of $ 3 \times 10^{16}$ GeV, corresponding to the upper
limit implied under the assumption that all fluctuations observed in COBE was due to tensor perturbations;
It is likely that the energy scale of inflation is significantly below that of the COBE limit.
As shown in  Figure~\ref{fig:sec},
there is another source of contribution to the curl component resulting 
from the fractional conversion of the gradient polarization 
via the weak gravitational lensing effect associated with projected potentials along the line of sight to the
last scattering surface\cite{ZalSel98}. 
While the lensing effect is generally at the level of few percent or below,
due to the fact that the polarization associated with density perturbations is several orders of magnitude larger than
that associated with curl modes, even an insignificant conversion at the level of sub percent or below
can be a significant source of confusion. Thus, the lensing-induced 
curl polarization introduces a noise from which gravitational waves must 
be distinguished in CMB polarization.

\begin{figure}[t]
\includegraphics[height=15pc,angle=-90]{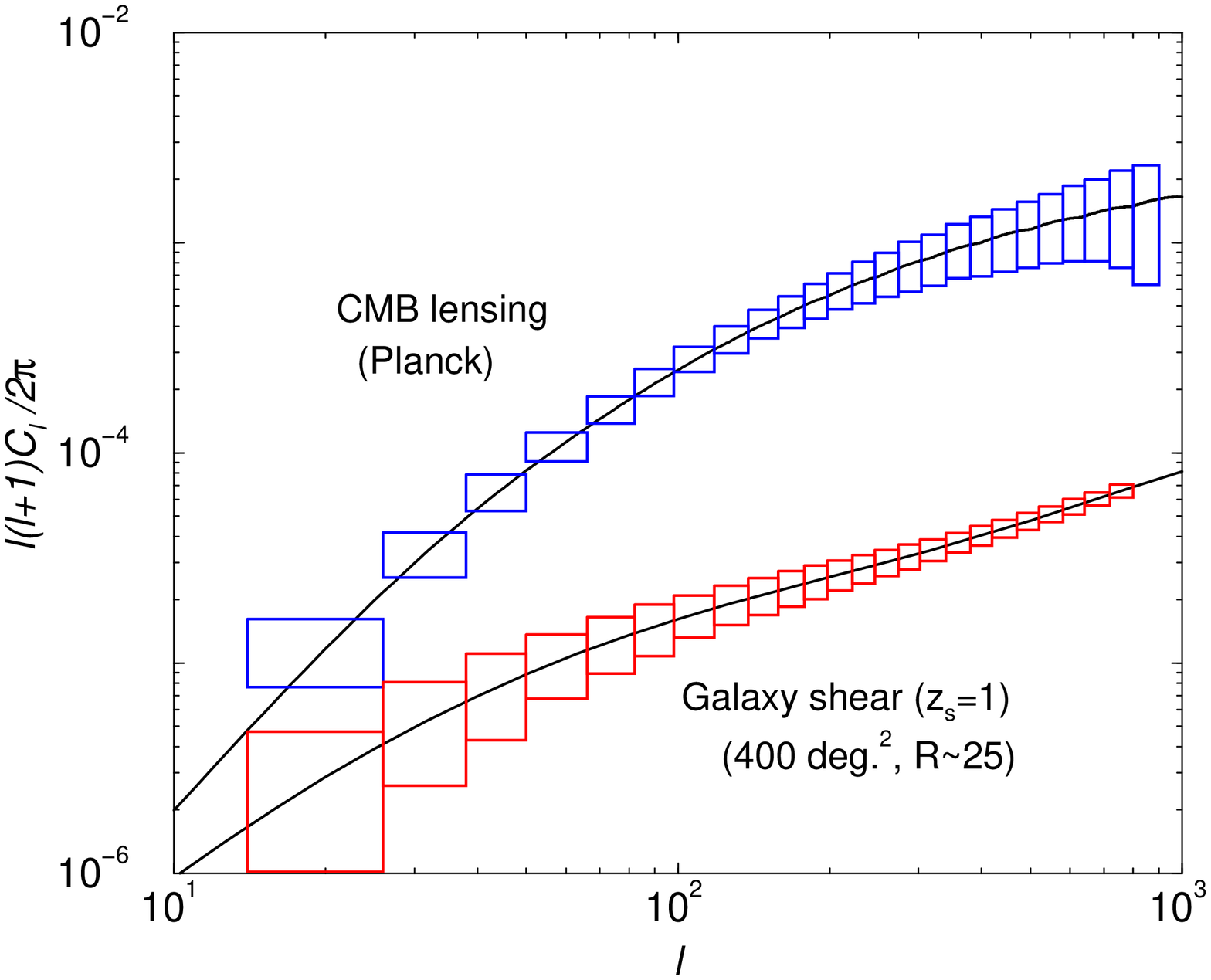} \includegraphics[height=15pc,angle=-90]{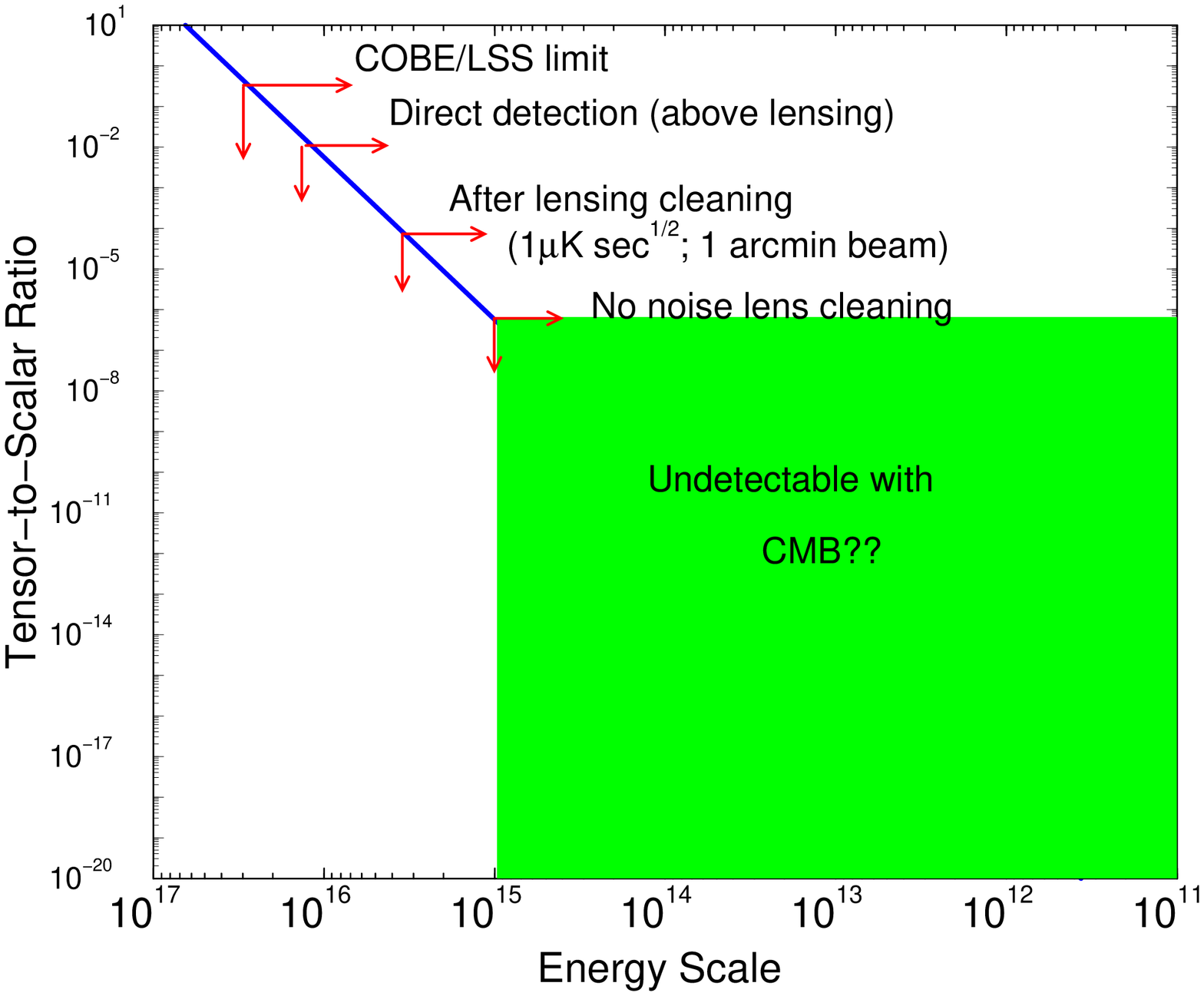} 
\caption{The reconstruction of lensing convergence with CMB temperature data from the Planck surveyor ({\it left}).
For comparison, we show expected errors on the convergence  measured from a galaxy lensing survey of 400 deg.$^2$ down
to a R-band magnitude of 25 and assuming all sources are at a redshift of 1. The amplitude difference between
the two curves shows the excess mass fluctuations beyond a redshift of 1 and responsible for the lensing of
CMB anisotropies. ({\it Right:}) Constraints on the inflaton potential, $V$, or the energy scale of inflation from
CMB polarization observations. The solid line shows the relation between $V$, and the 
amplitude of gravitational-wave contribution (with respect to amplitude of scalar perturbations).
The COBE data, under the assumption that all observed fluctuations are due to primordial
gravitational waves, provide an upper limit of $\sim 3 \times 10^{16}$ GeV.
The various limits correspond to what can be achieved with separation of cosmic shear confusion.
The ultimate limit for detection of inflationary gravitational waves corresponds to an energy scale
of $\sim 10^{15}$ GeV.}
\label{fig:limits}
\end{figure}

There are several ways to address the confusion and to reduce it such that a reliable detection of
inflationary gravitational waves is possible. If the projected matter distribution of large scale
structure, out to redshifts of $\sim$ 10, is known, one can clean the
polarization maps to partly reduce the confusion\cite{Coo02}. An alternative, and probably the best,
approach is to use CMB data themselves. The lensing signature in CMB temperature anisotropies can be
used to reconstruct the associated integrated-deflection angle\cite{Hu01} (Fig.~\ref{fig:limits}) 
which can then be used to
clean polarization maps. The technique can also be improved with polarization data as well\cite{HuOka01}.
While cleaning of the lensing signature from such techniques allow one to reduce the energy scale
that can be probed with polarization observations, it is unlikely that the separation can be
used to probe energy scales as low as $10^{12}$ GeV corresponding to certain super-symmetric models; 
with perfect all-sky maps,
current calculations on lensing extraction and separation techniques from CMB data indicate an ultimate limit of
around 10$^{15}$ GeV\cite{KnoSon02}. If inflation is associated with an energy scale below this limit, 
the gravitational wave signature will remain confused with cosmic shear and will require alternative
techniques to probe its presence.

As summarized in Figure~\ref{fig:sec}, the CMB temperature and polarization anisotropies contain distinct signatures of
the local universe and associated large scale structure\cite{Coo02a}. 
One such a well-known effect is the Sunyaev-Zel'dovich effect
associated with inverse-Compton scattering of CMB photons via hot electrons in galaxy clusters\cite{SunZel80}. 
The effect has now
been imaged to massive clusters where CMB temperature change is of order few 100s of micro Kelvin or more\cite{Caretal96}
and may have been detected statistically as a secondary anisotropy signal in recent CBI data \cite{Masetal02}.
The extent to which the SZ signal is present can be confirmed through multifrequency measurements and the upcoming
ACBAR analysis will be useful for this purpose.

Due to the behavior of inverse-Compton scattering, the SZ effect has a distinct frequency dependence and
allows its separation from fluctuations associated with thermal CMB and other foregrounds\cite{Cooetal00}. 
The SZ spectrum has been previously considered for a measurement of
the peculiar velocity of galaxy clusters, since an additional secondary effect in clusters involves the
Compton-scattering via electrons moving with respect to the rest frame of CMB; 
the latter Doppler effect has the same frequency dependence as the CMB
black body spectrum. The separation of thermal and kinetic SZ effects based on limited multifrequency data,
 is, in general, hard and only upper limits
are currently available for the peculiar velocity of several clusters\cite{Holetal97}.

\begin{figure}[t]
\begin{center}
\includegraphics[height=13pc,angle=0]{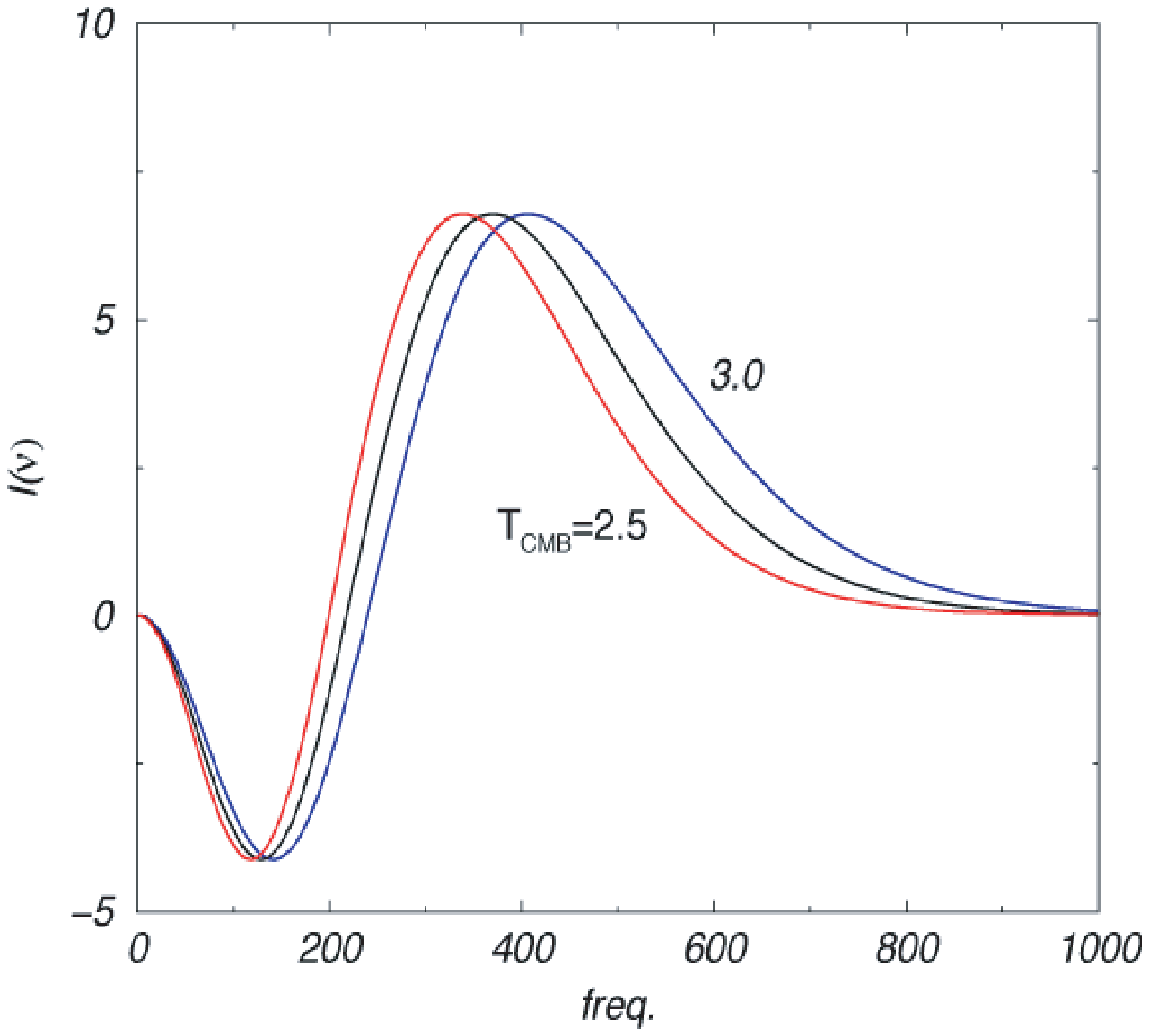} \includegraphics[height=11pc,angle=0]{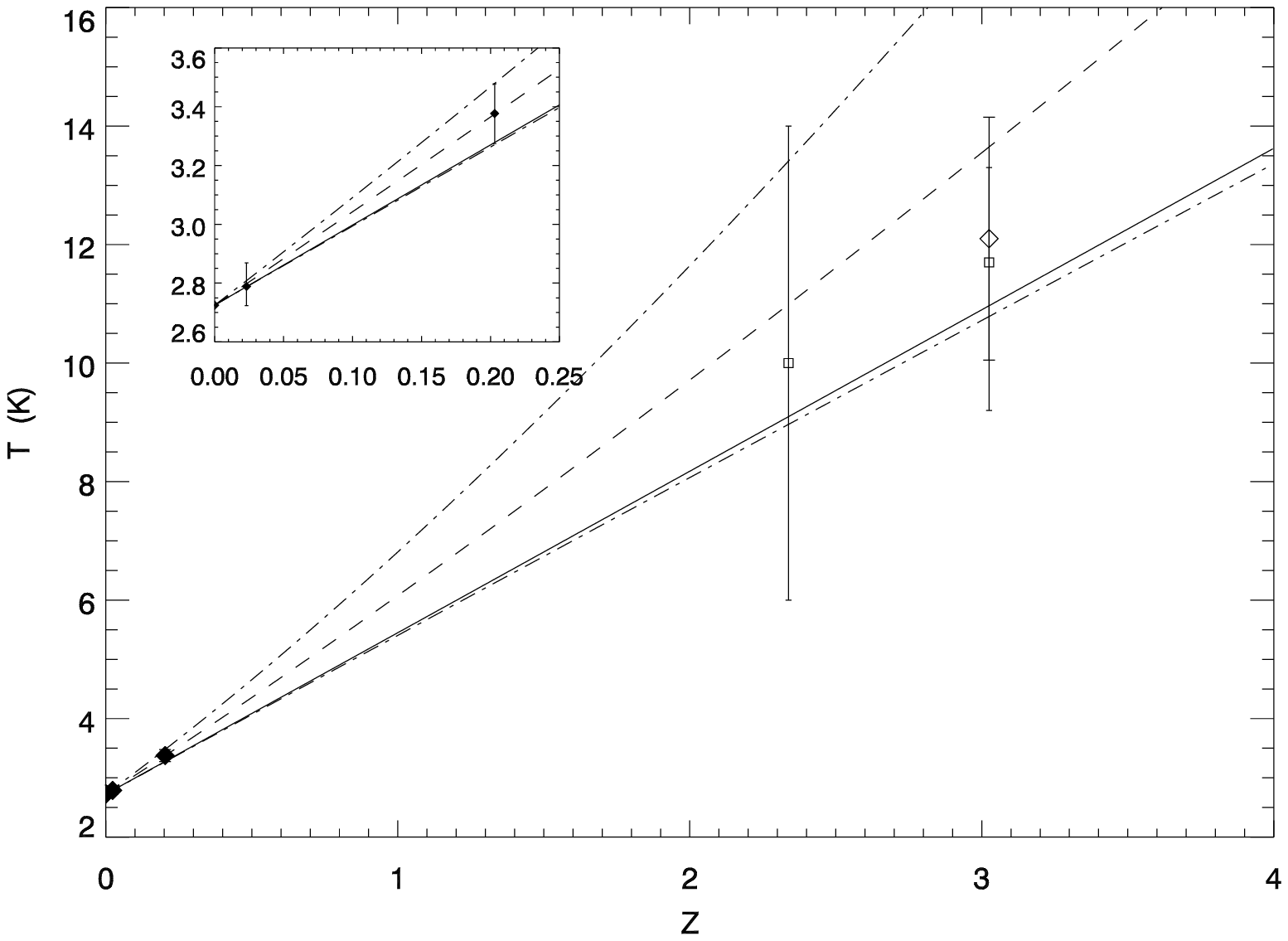} 
\end{center}
\caption{{\it Left:} The SZ spectrum  as a function of the assumed CMB temperature.
The multifrequency measurements of the SZ effect allows one to infer the absolute temperature of CMB at 
the redshift of the cluster. {\it Right:} The CMB temperature as a function of the redshift. The 
inset shows a magnified view of two estimates based on SZ data of two clusters.
The other estimates, at redshifts greater than 2, comes from molecular transition line measurements
and the local measurement comes from the COBE/FIRAS experiment. We refer the reader to Battistelli et al. (2002) for
details.}
\label{fig:sz}
\end{figure}

The multifrequency observations of the SZ effect, however, is useful for other purposes. A potentially interesting and 
unique application was recently discussed in Battistelli et al.\cite{Batetal02} involving a 
measurement of the CMB temperature from the SZ spectrum (Figure~\ref{fig:sz}). 
During the Planck-era, multifrequency observations of the SZ effect can be utilized
for a high precision  test of the expected $T_{\rm CMB}(z)=T_{\rm CMB}(1+z)$ 
dependence of the background temperature evolution in standard cosmological models. 
While mostly ignored in the literature, simple applications such as the one
involving CMB temperature measurement with the SZ spectrum are useful for variety of purposes
and should be considered in future even for ground based multifrequency SZ experiments.

While detailed attention has been made with respect to secondary effects in CMB temperature and their
cosmological and astrophysical uses, with increasing experimental developments, the secondary
polarization signals will also become important. We briefly discussed the role played by
gravitational lensing when trying to extract information related to the early universe, mainly inflation,
from CMB polarization data. An additional source of confusion is expected from polarization signals associated
with scattering of CMB temperature quadrupole via electrons in galaxy clusters\cite{Bauetal02}. 
As shown in Fig.~\ref{fig:sec},
the cluster contributions to polarization power spectra are at least two to three orders of magnitude below the
lensing confusion. If lensing effect is removed to such a high precision, cluster confusions may become
important. 

On the other hand, observations of galaxy cluster polarization are useful for several unique studies. One such
possibility is related to extracting some information related to  the CMB temperature quadrupole\cite{CooBau02}.
Since at low redshifts, the temperature quadrupole contains the signature of time evolving potentials,
mainly the integrated Sachs-Wolfe effect \cite{SacWol67} (ISW), with a sample of clusters widely distributed over
the redshift space, one can essentially measure the redshift evolution of the ISW contribution. To carry out such a study 
reliably, multifrequency polarization observations are required. This is due to the fact that an additional
polarization contribution in clusters comes from the kinematic quadrupole associated with the transverse motion. 
Fortunately, polarization contributions related to kinematic and temperature
quadrupoles have different frequency dependences such that the two can be separated with their spectral information.

While polarization information related to the kinematic quadrupole captures the transverse velocities of clusters
on the sky, the polarization associated with the temperature quadrupole provides several new opportunities for both
cosmological and astrophysical studies. Since the ISW effect depends strongly on details related to the dark energy, a 
reliable reconstruction of the ISW quadrupole evolution can eventually be used as a probe of dark energy properties.
Similarly, with the polarization information distributed in redshift space, one can reconsider various cross-correlation
studies to understand astrophysical connections between the evolving large scale structure and the ISW effect.

\begin{figure}[t]
\begin{center}
\includegraphics[height=20pc,angle=-90]{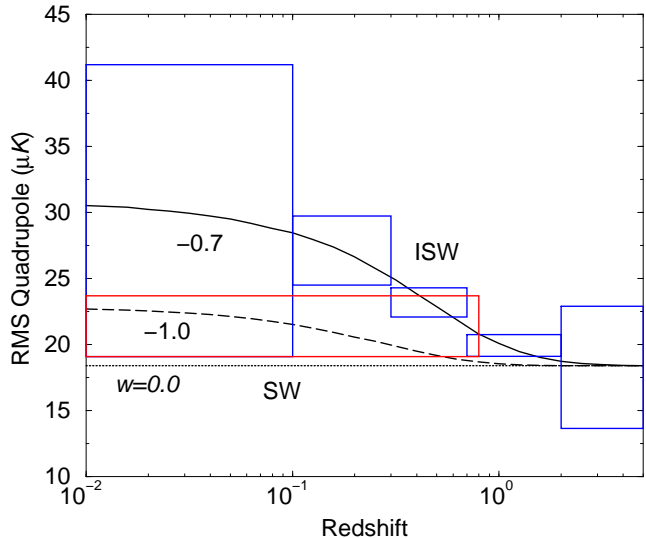} 
\end{center}
\caption{The rms temperature quadrupole as a function of redshift. The error
bars on the top curve illustrate the expected uncertainties in the reconstruction
of rms temperature quadrupole from followup multi-frequency polarization
observations of galaxy clusters with a catalog down to  10$^{14}$ M$_{\odot}$ and covering
10,000 deg$^2$. The large single error on the middle curve illustrate the expected uncertainty with the whole-sky
Planck cluster catalog down to a mass limit of $5 \times 10^{14}$ M$_{\odot}$
and using Planck polarization observations of same clusters.}
\label{fig:pol}
\end{figure}

\section*{Acknowledgements}
The author thanks Daniel Baumann, Marc Kamionkowski, Mike Kesden,
and Kris Sigurdson for collaborative work and useful discussions. 
This work is supported at Caltech by a senior research fellowship from the Sherman Fairchild
foundation and grants from the Department of Energy and the American Astronomical Society.

\end{document}